\documentclass[a4paper,11pt]{article}
\pdfoutput=1 

\usepackage{jheppub} 
                     
\usepackage{amsmath,amssymb,amsthm,amscd,graphicx}
\usepackage{psfrag}
\input epsf.sty

\usepackage{color}

\theoremstyle{definition}

\newcommand{\fad}{\operatorname{\Phi}_{\mathsf{b}}}

\newcommand{\kk}{K(k)}
\newcommand{\kkp}{K'(k)}

\def\IZ{{\mathbb Z}}
\def\IR{{\mathbb R}}

\def\IP{{\mathbb P}}

\def\IF{{\mathbb F}}

\newcommand{\tr}{{\rm Tr}}
\newcommand{\re}{{\rm e}}
\newcommand{\ri}{{\rm i}}
\newcommand{\rd}{{\rm d}}

\newcommand{\Li}{\mathop{\rm Li}\nolimits}

\newcommand{\be}{\begin{equation}}
\newcommand{\ee}{\end{equation}}
\newcommand{\ba}{\begin{aligned}}
\newcommand{\ea}{\end{aligned}}
\newcommand{\ben}{\begin{eqnarray}\displaystyle}
\newcommand{\een}{\end{eqnarray}}

\newdimen\tableauside\tableauside=1.0ex
\newdimen\tableaurule\tableaurule=0.4pt
\newdimen\tableaustep
\def\phantomhrule#1{\hbox{\vbox to0pt{\hrule height\tableaurule width#1\vss}}}
\def\phantomvrule#1{\vbox{\hbox to0pt{\vrule width\tableaurule height#1\hss}}}
\def\sqr{\vbox{%
  \phantomhrule\tableaustep
  \hbox{\phantomvrule\tableaustep\kern\tableaustep\phantomvrule\tableaustep}%
  \hbox{\vbox{\phantomhrule\tableauside}\kern-\tableaurule}}}
\def\squares#1{\hbox{\count0=#1\noindent\loop\sqr
  \advance\count0 by-1 \ifnum\count0>0\repeat}}
\def\tableau#1{\vcenter{\offinterlineskip
  \tableaustep=\tableauside\advance\tableaustep by-\tableaurule
  \kern\normallineskip\hbox
    {\kern\normallineskip\vbox
      {\gettableau#1 0 }%
     \kern\normallineskip\kern\tableaurule}%
  \kern\normallineskip\kern\tableaurule}}
\def\gettableau#1{\ifnum#1=0\let\next=\null\else
\squares{#1}\let\next=\gettableau\fi\next}

\title{\boldmath Seiberg--Witten theory as a Fermi gas }
\author{ Giulio Bonelli$^a$, Alba Grassi$^b$ and Alessandro Tanzini$^a$}

\affiliation{
$^a$ International School of Advanced Studies (SISSA), \\
via Bonomea 265, 34136 Trieste, Italy and INFN, Sezione di Trieste\\
\\
$^b$International Center for Theoretical Physics,\\
 ICTP, Strada Costiera 11, Trieste 34151,Italy \\}

\emailAdd{bonelli@sissa.it, agrassi@ictp.it, tanzini@sissa.it}
\preprint{
\begin{flushright}
SISSA  15/2016/FISI-MATE \\
\end{flushright}
}

\abstract{
 We explore a new connection between Seiberg--Witten theory and quantum statistical systems   by relating 
 the dual partition function of  $SU(2)$ Super Yang-Mills  theory in a self--dual $\Omega$--background  to the spectral determinant of an ideal Fermi gas.  We show that the spectrum of this gas  is encoded in the zeroes of the Painlev\'e ${\rm III}_3$ $\tau$ function. In addition we find that the Nekrasov partition function on this background  can be expressed as an $O(2)$ matrix model.
 Our construction  arises as a four-dimensional limit of a recently proposed conjecture relating topological strings and spectral theory. In this limit, we provide a mathematical proof of the conjecture  for the local  $\IP^1 \times \IP^1$ geometry. 
 }

\begin{document}
\maketitle

\flushbottom

\section{Introduction}
Supersymmetric partition functions of gauge theories in four dimensions can be regarded as a new class of special functions which permeates many subjects in physics and mathematics,
such as classical \cite{Gorsky:1995zq,Martinec:1995by,Donagi:1995cf} and quantum integrable systems \cite{ns,franco1, franco2}, two-dimensional conformal field theories \cite{agt} and quantum hydrodynamics \cite{Ntalk,Otalk,Alba:2010qc,Bonelli:2014iza,Bonelli:2015kpa,ks1}, differential \cite{Nekrasov:2003viNO,Gottsche:2006tn,ronzani} and enumerative geometry \cite{kkv}. 

In this paper we explore a new aspect of this web of connections, which concerns quantum statistical systems. More precisely, we 
show that the dual partition function of pure $\mathcal{N}=2$  $SU(2)$ Super Yang-Mills in four dimensions in a self--dual $\Omega$--background  is the spectral determinant of a quantum system given by an ideal Fermi gas.
This in turn is also related to a statistical model describing self-avoiding polymers in two-dimensions.

The core of our work relies on the merging of two topics.  
The first one is the connection between four-dimensional $\mathcal{N}=2$ gauge theories and Painlev\'e transcendants found in \cite{ilt1,ilt,gil1,gil,bes,ilte}
and the other is a recently proposed conjecture relating topological strings and spectral theory \cite{ghm}.

The main result of \cite{ilt1,ilt,gil1,gil,bes,ilte} is that the Nekrasov--Okounkov (NO) partition function of  $SU(2)$ four-dimensional gauge theory calculates the $\tau$ function of a corresponding Painlev\'e equation. More precisely, for the pure ${\mathcal N}=2$ gauge theory it has been shown that the NO partition function, defined in terms of the standard Nekrasov partition function as 
\be\label{NO}Z^{\rm NO}(\eta,a/\epsilon,T)=\sum_{n \in \IZ} e^{4\pi \ri  n \eta}Z^{\rm Nek}(\epsilon,a+n\epsilon,T)\, ,\ee
computes  the general $\tau $ function of the Painlev\'e $\rm{III}_3$ equation. The gauge theory parameters $a$ and $\eta$ are related to the initial conditions of the specific solution.
A thorough discussion of this relation including new long distance expansions of solutions to Painlev\'e equations and their relation with Argyres--Douglas points
will be presented elsewhere \cite{blmst}.

On the other hand,
in the conjecture of \cite{ghm},  one associates to any toric Calabi--Yau threefold X
a trace class operator $\mathsf{\rho}_X$ whose 
spectral determinant 
computes the non--perturbative topological string partition function. More precisely one has
\be \label{conj} \det \left(1+\kappa \rho_X \right)= \sum_{n \in \IZ}\re^{\mathsf{J}(\mu+2\pi \ri n)} \qquad \kappa=\re^{\mu} \, \ee
where $\mathsf J$ is the non--perturbative topological string free energy (or grand  potential).

It is well known that  a class of $\mathcal{N}=2$ gauge theories in four dimensions can be engineered by using  toric Calabi--Yau threefolds \cite{kkv,ikp}.
For instance, by taking the so--called  geometric engineering limit of local $\IP^1 \times \IP^1 $ one generates the $\mathcal{N}=2$ supersymmetric  $SU(2)$ gauge theory out of string theory. More precisely, different phases of the four-dimensional theory in the gravitational background parameters can be obtained 
by different scaling limits of the refined topological string. In \cite{hm} the authors specified the conjecture of \cite{ghm} in the  four-dimensional limit corresponding to the Nekrasov-Shatashvili (NS) regime ($\epsilon_2=0$) and 
 the resulting spectral problem was found to coincide with the quantum Toda system found in \cite{ns}.
 
In this paper we study a different regime in the $\Omega$-background parameters of the four-dimensional gauge theory, namely $\epsilon_1=-\epsilon_2=\epsilon$.
Moreover, we scale the K\"ahler parameters of the Calabi-Yau in terms of $\epsilon$,  therefore implementing a limit which is slightly different from the usual geometric engineering one,
see Section \ref{sp} for details.
In this limit we find that the conjecture \eqref{conj} connects to the NO partition function \eqref{NO} with $\eta=0$.

Our result can be interpreted in a twofold way. 
On one side it is a proof of  the conjecture \eqref{conj} in the four-dimensional limit. 
On the other, it is a new connection between Seiberg-Witten theory on self-dual background and the spectral theory of an ideal Fermi gas.

This paper is organized as follows.
In the next section we recall the content of the conjecture of \cite{ghm} in a form suitable for the subsequent manipulations.

In Section \ref{sp} we give a proof for the conjecture of \cite{ghm} in the  four-dimensional limit  by using 
the well known results on spectral determinants \cite{zamo} together with  
the recent developments in the context of  Painlev\'e  equations \cite{ilt1,ilt,gil1,gil,bes,ilte}. 
More precisely we  show that both the l.h.s~and the r.h.s~of \eqref{conj} satisfy the Painlev\'e $\rm{III}_3$ equation in the $\tau$ form
with the same initial conditions. 

In Section \ref{gauge} we analyze our results from the four-dimensional gauge theory perspective. 
We show that the four-dimensional $SU(2)$ Nekrasov--Okounkov partition function can be expressed by using the grand  partition function of an ideal Fermi gas. This gas is characterized by the following density operator
 \be \label{opri} \rho= \re^{-{4 T^{1/4} } \cosh(\hat x)}{4 \pi \over \left(\re^{\hat p/2}+ \re^{-\hat p/2} \right)}\re^{-{4 T^{1/4} } \cosh(\hat x)}, \qquad [\hat x, \hat p]= 2 \pi \ri  .\ee
This means that we have a one-dimensional  ideal Fermi gas in an external confining potential
\be V(x)={8 T^{1/4} } \cosh( x).\ee
In this way we provide a Fermi gas picture of gravitational corrections to the Seiberg--Witten theory.
Notice that this picture is  different from the one of \cite{no2}, where the authors consider free chiral fermions lying on the Seiberg--Witten curve.

As a consequence of our computation we can write the Nekrasov partition function as an $O(2)$ matrix model \cite{Kostovo2, kos39}
 which reads
 \be \label{mmsw}Z_{\rm 4d}(N,T)={1\over N!} \int \prod_{i=1}^N  \frac{\rd z_i}{z_i} \re^{- 4 T^{1/4} \sum_{i=1}^N \left(z_i +{1\over z_i}\right)} \prod_{i < j} \frac{(z_i-z_j)^2}{(z_i+z_j)^2}.\ee
 The 't Hooft expansion of this model reproduces the standard perturbative expansion of the gauge theory in a self--dual $\Omega$--background in the magnetic frame. 
 As we will see in Section \ref{gauge2} the 't Hooft parameter $N T^{-1/4}$ of the model has an explicit meaning in the gauge theory   as the dual period of the Seiberg--Witten curve.
 Moreover  \eqref{mmsw} is manifestly convergent, well-defined also for finite $N $ and it leads  to an exact expression for the Nekrasov partition function which resums the instanton expansion. 
  We would like also to stress that our model  is derived without using  Nekrasov's results but simply by quantizing the mirror curve and by taking a particular limit.  Therefore the fact that \eqref{mmsw} reproduces Nekrasov's  partition function is a highly non--trivial fact.

\section{The  spectral theory/topological string conjecture}\label{intro1}
In this section we briefly review the conjecture of \cite{ghm}. For a pedagogical review see \cite{mmnew}.

In \cite{ghm}, based on previous works \cite{km,hmmo,hw,cgm,ghmabjm}, it has been conjectured that there is a precise and exact duality between spectral theory and enumerative geometry. 
In this construction one associates a non--pertubative  trace class operator $ \rho_X$ to any toric Calabi--Yau $X$ whose mirror curve has genus one \footnote{The generalization to higher genus mirror curves was subsequently worked out in \cite{cgm2}. }. This operator has a discrete spectrum, denoted by $\re^{-E_n}$, which can be organized into a  spectral determinant:
\be \label{sd1} \Xi_X(\kappa)=\det(1+\kappa \rho_X)=\prod_{n\geq 0} \left(1+\kappa \re^{-E_n}\right). \ee
 We also refer to \eqref{sd1}  as Fredholm determinant.

In this paper the relevant geometry is the anti-canonical bundle of $\IP^1\times \IP^1$ and the corresponding mirror curve is
\be \label{mirr} \re^{ x}+\re^{ p}+\re^{- p}+m_{\mathbb{F}_0}\re^{- x}-\kappa=0,\ee
where $\kappa=\re^{\mu}$ is the so--called complex modulus of $\IP^1\times \IP^1$  while $m_{\mathbb{F}_0}$ is the so--called mass parameter.
In the following we will  often use the "renormalized" mass parameter introduced in  \cite{kmz}
\be \label{smallm} \log m={2\pi \over \hbar}\log m_{\mathbb{F}_0}. \ee
A detailed description of this geometry and its enumerative content can be found in  \cite{bz, hkp,hkrs}.
The trace class operator arising in the quantization of the mirror curve \eqref{mirr} is  
\be \label{rhop1p1}\rho_{\IP^1 \times \IP^1}=\left( \re^{\hat x}+\re^{\hat p}+\re^{-\hat p}+m_{\mathbb{F}_0}\re^{-\hat x}\right)^{-1}, \quad [\hat x,\hat p]=\ri \hbar, \quad \hbar, m_{\mathbb{F}_0}>0. \ee
The conjecture of \cite{ghm} states that the spectral determinant  \eqref{sd1}  can be written explicitly in terms of (refined) GV invariants of $X$ as
\be \label{sd2}\Xi_X(\kappa,\hbar)=\sum_{n\in \IZ}\re^{\mathsf J(\mu+2\pi n \ri,\hbar)} , \quad \kappa=\re^{\mu}\ee
 where the grand potential, $\mathsf J(\mu, \hbar)$ can be expressed in terms of enumerative invariants  of $X$. 
This quantity was first introduced in the context of ABJM in \cite{mp} and subsequently studied in a series of works  \cite{hmo, hmo2,hmo3,km,cm,hmmo,ghm,gkmr,hatsuda}.
More precisely we have
 \be\label{Jex} \mathsf{J}({\mu}, \hbar) = \mathsf{J}^{\rm WKB}({\mu},\hbar)+ \mathsf{J}^{\rm WS}({\mu}, \hbar).
\ee
  The WKB part is given by
  \be
\label{jm2}
\mathsf{J}^{\rm WKB}({\mu}, \hbar)= {t_i( \mu, \hbar) \over 2 \pi}   {\partial F^{\rm NS}({ \bf t}(\mu, \hbar), \hbar) \over \partial t_i} 
+{\hbar^2 \over 2 \pi} {\partial \over \partial \hbar} \left(  {F^{\rm NS}({ \bf t}(\mu, \hbar), \hbar) \over \hbar} \right) + {2 \pi \over \hbar} b_i {t_i}(\mu, \hbar) + A({ m}, \hbar),
\ee
where $F^{\rm NS}$ is  the Nekrasov--Shatashvili free energy \cite{ns},  $A({ m}, \hbar)$ the so--called constant map contribution \cite{bcov} and $b_i$ are  real numbers related to the genus one free energy (see \cite{cgm2} for a detailed explanation). 
We  denote by
 $t_i(\mu, \hbar)$  the quantum K\"ahler  parameters of $X$ and by $\mu$ the complex modulus of the mirror geometry.
In addition $\mathsf J(\mu, \hbar)$ can depend on a set of mass parameters. However,  for sake of notation, we do not explicitly write  mass dependence. 
The "worldsheet" part in \eqref{Jex} is given by
\be \mathsf{J}^{\rm WS}_X({\mu},  \hbar)=F^{\rm GV}\left( {2 \pi \over \hbar}{\bf t}( \mu, \hbar)+ \pi \ri { B} , {4 \pi^2 \over \hbar} \right),\ee
 where $F^{\rm GV}$ denotes the (unrefined) Gopakumar-Vafa \cite{gv} free energy and $B$ is the so--called $B$ field. 
 Both  $F^{\rm GV}$ and $F^{\rm NS}$ are  expressed explicitly in terms of (refined) enumerative invariants \cite{gv,ckk,no} as explained in appendix \ref{ap1} .
 
 For the local $\IP^1 \times \IP^1$ geometry we have \cite{ghm}
 \be t_1( \mu, \hbar)=t ( \mu, \hbar), \quad t_2=t ( \mu, \hbar)-\log m_{\IF_0}\ee
  where $t( \mu, \hbar)$ is the quantum A-period of the curve \eqref{mirr}, it is computed in appendix \ref{ap2}.
 Moreover the B field for this geometry is  $B=2$ and the $b_i$ coefficients in \eqref{jm2} are
 \be {\bf b}=\{1/24,1/24\} .\ee 
The  explicit expression for  the constant map contribution $A(m,\hbar)$ of local $\IP^1 \times \IP^1$ has been computed in  \cite{ho2}
and it reads
\be \label{cm} A(m,\hbar)= {\hbar^2 \over (4 \pi^2)^2}\left[\frac{\log ^3(m)}{24}+\frac{\pi^2  \log (m)}{6 }\right]+A_c\left({\hbar\over \pi}\right)-F_{\rm CS}\left({\hbar\over \pi}, {2\pi \hbar  +\ri \hbar  \log m\over 4 \pi^2}\right),\ee
where
\be
A_{\rm c}(k)= \frac{2\zeta(3)}{\pi^2 k}\left(1-\frac{k^3}{16}\right)
+\frac{k^2}{\pi^2} \int_0^\infty \frac{x}{\re^{k x}-1}\log(1-\re^{-2x})\rd x,
\ee
and $F_{\rm CS}(k,M)$ is the $U(M)$ Chern--Simons (CS) free energy.
In the context of CS theory  $M$ and $k$ are integers. However in
the expression for the constant map \eqref{cm} they are complex. Therefore we need to analytically continue the CS partition function. This analytic continuation is not unique, however the spectral problem associated to local $\IP^1 \times \IP^1$  fixes this ambiguity \cite{gkmr}
and the relevant analytic continuation turns out to be the one proposed in \cite{ho2}, namely
\be\label{csm} \ba 
F_\text{CS}\left({\hbar \over \pi},{2\pi \hbar  +\ri \hbar  \log m\over 4 \pi^2}\right)&=\frac{\hbar^2}{8 \pi^4}\left(\Li_3(-m^{1/2})+\Li_3(-m^{-1/2})-2 \zeta(3)\right)\\
&+\int_0^\infty \rd x \frac{x}{\re^{2\pi x}-1}
\log \left(\frac{4 \sinh^2 \frac{\pi^2 x}{\hbar} }{4  \sinh^2 \frac{\pi^2 x}{\hbar}+(m^{1/4}+m^{-1/4})^2} \right).
\ea \ee
 Even though the conjecture of \cite{ghm} has been successfully tested in many examples \cite{ghm,kama,mz,kmz,ho2,gkmr,oz,hw, wzh}, a proof is still missing.
 In this paper we perform a first step in this direction by proving the conjecture in a four-dimensional limit of local $\IP^1 \times \IP^1$.
More precisely we  show that, in this limit, the spectral determinant \eqref{sd1} and its conjectural expression \eqref{sd2} both fulfill the Painlev\'e $\rm{III}_3$ equation with the same asymptotic behavior.

\section{The four--dimensional limit} \label{sp}
In this section we prove the conjecture \cite{ghm} in a particular double-scaling limit of  local $\IP^1\times \IP^1$.
 This double-scaling limit is defined as follows.
Let us denote by
\be  \label{qqf} Q=\re^{-{2\pi \over \hbar}t(\mu, \hbar)}, \qquad Q_f=m Q, \qquad m>0, \quad \hbar>0. \ee
We  parametrize the above quantities by
\be \ba  \label{ftl} q=\re^{ {4\pi^2 \ri \over \hbar}}&= \exp (-\beta  \epsilon ), \quad Q= (\beta  \epsilon)^4 T, \quad Q_f= \exp (-2 \sigma \epsilon \beta ), \\
\\
& \quad \beta > 0, \quad \ri \epsilon >0, \quad T>0 \quad  \sigma \epsilon \in \IR/\{0\}. \ea\ee
The four-dimensional limit we are going to study  consists in taking $\beta \to 0$ while keeping the other parameters fixed. In particular this means that
\be m, \hbar \to \infty, \ee
therefore we are probing the non--perturbative regime of \cite{ghm}. 
{In this limit one of the two K\"ahler classes of $\IP^1 \times \IP^1 $ goes to infinity while the other is keep fixed. }In \cite{ghm} $m, \hbar$  are positive numbers, therefore 
$\ri \epsilon$ and $\ri \sigma $ are  real and different from zero.  

Notice  that the scaling  \eqref{ftl} is slightly different from the typical four-dimensional scaling of geometric engineering \cite{ikp, kkv} where one defines
\be \label{sc2} Q=\re^{-t(\mu, \hbar)}=(\beta  \Lambda /2)^4, \qquad Q_f=m_{\mathbb{F}_0} Q= \exp (-2 a \beta ), \qquad \hbar=\beta \epsilon \ee
and takes $\beta \to 0$ i.e.
\be m_{\mathbb{F}_0} \to \infty, \qquad \hbar \to 0.\ee 
In the scaling limit \eqref{sc2} the conjecture of \cite{ghm} makes contact with the  $SU(2)$ quantum Toda as shown in \cite{hm}. 
However to make contact with Painlev\'e $\rm{III}_3$ equation we have  to consider the scaling \eqref{ftl}.

In the following we compute the four-dimensional limit \eqref{ftl} of the spectral determinant \eqref{sd1}, see Subsection \ref{4dope}, and of its conjectural expression \eqref{sd2}, see Subsection \ref{topstring}, and  we show that 
they both satisfy Painlev\'e $\rm{III}_3$ equation in the $\tau$ form with the same initial conditions.  

\subsection{The topological string computation}\label{topstring}
In this section we focus on  the four-dimensional limit of  \eqref{sd2} for the local $\IP^1 \times \IP^1 $ geometry.

The standard and NS free energies determining the grand potential of this geometry are computed in appendix \ref{ap1}. 
By using these results we can write the grand  potential \eqref{Jex} as  
\be \ba \label{j4d} \mathsf{J}_ {\IP^1\times \IP^1}({\mu},  \hbar)= &\mathsf J^{\rm inst}(Q_f, Q, \hbar)+\mathsf J^{\rm one loop}(Q_f, \hbar)+ A(m,\hbar)+ \mathsf P(t(\mu, \hbar), m, \hbar) +\mathsf{J}^{\rm M}(Q_f, Q, \hbar).~ \\
\ea \ee
 We denote by $\mathsf J^{\rm inst}$ the instantons part of the  Gopakumar--Vafa free energy \eqref{gvp1p1}, namely
 \be \ba &\mathsf J^{\rm inst}(Q_f, Q, \hbar)= \frac{2 q Q}{(q-1)^2 (Q_f-1)^2}+\\
&\frac{q^2 Q^2 \left(q^2 Q_f^4+q^2+4 q (q+1)^2 Q_f^3-2 (q (q+1) (q (q+3)+4)+1) Q_f^2+4 q (q+1)^2 Q_f\right)}{(q-1)^2 (q+1)^2 (Q_f-1)^4 (q-Q_f)^2 (q Q_f-1)^2}+\mathcal{O}(Q^3), \ea\ee
where $q=\re^{4 \pi^2\ri \over \hbar }$. The one-loop part is
 \be \ba  \label{f1lr}\mathsf J^{\rm one loop}(Q_f, \hbar)=&  -\sum_{m\geq 1}\frac{ Q_f^m}{2m \sin \left(\frac{2 \pi^2 m}{\hbar}\right)^2} +\sum_{m\geq 1}\frac{Q_f^{\frac{\hbar m}{2\pi}}}{4 \pi   m^2}2 \cot \left(\frac{\hbar m}{2}\right)\left[1-{\hbar  m \over 2\pi }\log (Q_f)\right] \\
& +\sum_{m\geq 1}Q_f^{\frac{\hbar m}{2\pi}}\csc ^2\left(\frac{\hbar m}{2}\right){  \hbar \over  4 \pi m} .\ea \ee
In our notation $\mathsf J^{\rm one loop}(Q_f, \hbar)$ contains the one-loop contribution  both of standard and NS free energies.
 By using the techniques of \cite{ho2,hm}, illustrated in appendix \ref{ap3}, we can write \eqref{f1lr}  in closed form as
  \be \label{intol} \mathsf J^{\rm one loop}(Q_f, \hbar)= - \frac{\hbar^2}{8 \pi ^4} \text{Li}_3(Q_f) 
+2 {\rm Re} \int_0^{\infty\re^{\ri 0}} \rd x\frac{x}{\re^{2\pi x}-1}\log(1+Q_f^2-2Q_f\cosh {4 \pi^2x \over \hbar}).\ee
Notice that the one--loop term in the gauge theory typically requires a certain regularization scheme and there exists a scheme in which one recovers the one--loop term of the topological vertex \cite{ikv,hiv}. This is the same  scheme used in the context of AGT correspondence (see appendix B.2 of \cite{agt}). As  was pointed out there, this  is different from the scheme used in the NO partition function \cite{no2}.
Remarkably, the grand potential \eqref{Jex}  automatically implements the correct scheme to make contact with the NO partition function and the Painlev\'e $\rm{III}_3$ equation. Similarly the analysis of \cite{hm} shows that \eqref{Jex} also implements the correct scheme  in the context of $SU(2)$ Toda.

In \eqref{j4d} we denote by $\mathsf P(t(\mu, \hbar), m, \hbar) $ the polynomial part of the grand potential, which reads
\be \mathsf P(t(\mu, \hbar), m, \hbar)= -\frac{\log (m) t(\mu ,\hbar )^2}{16 \pi ^2}+\frac{t(\mu ,\hbar )^3}{12 \pi  \hbar }-\frac{\hbar  t(\mu ,\hbar )}{24 \pi }+\frac{\pi  t(\mu ,\hbar )}{6 \hbar }-\frac{\log (m)}{24}.  \ee
The last term  in \eqref{j4d} 
\be \label{wan} \mathsf{J}^{\rm M}(Q_f, Q, \hbar)\ee
denotes the instanton part of the WKB grand potential \eqref{jm2} which is completely determined by the instanton part of the NS free energy  \eqref{nsintsr}.  This term will not be important here since it vanishes in the four-dimensional limit \eqref{ftl} where we have
\be \re^{-t_1(\mu, \hbar)}=\left(\beta ^4 T \epsilon ^4\right)^{\frac{2  \pi }{\beta \ri  \epsilon }}, \qquad   \re^{-t_2(\mu, \hbar)}= e^{4 \ri \pi  \sigma }. \ee
It is straightforward to see that \eqref{nsintsr}, and as a consequence the instanton part of the WKB grand potential \eqref{wan},  vanishes as we take $\beta \to 0$.  Therefore we are studying precisely the opposite regime with respect to the one considered in the standard  geometric engineering limit \eqref{sc2} and in the context of Toda systems \cite{hm,ns}.

To compute the spectral determinant \eqref{sd2} we have to perform the shift
\be \mu \to \mu -2 \pi \ri n.\ee
 By using the expression for the quantum A-period given in  appendix \ref{ap2} we have
 \be t(\mu -2\pi n \ri, \hbar)=t(\mu , \hbar)-4 \pi \ri n.\ee
Similarly  the quantities in \eqref{qqf}  shift according to
\be \label{sQQf} Q\to Qq^{2n} \qquad Q_f\to Q_fq^{2n} , \qquad q=\re^{4 \pi^2\ri \over \hbar }.\ee
We are now going to compute the four-dimensional limit of each quantity in \eqref{j4d}.

\subsubsection{Instanton contribution }
By using the refined topological vertex \cite{ikv}, together with \eqref{ftl} and \eqref{sQQf}
we have
\be \label{4dins}  \mathsf J^{\rm inst}(q^{2n}Q_f, q^{2n}Q, \hbar)\quad \xrightarrow{4D} \quad\sum_{k\geq 1}  c_k(\sigma,n) T^{k} ,\ee
where  $c_k(\sigma,n)$ are the well-known coefficients of the  Nekrasov partition function  \cite{n,Bruzzo:2002xf,Flume:2002az}. 
 The first few coefficients read
\be \label{ccoeff}\ba   c_1(\sigma,n)=&\frac{1}{2  (\sigma +n  )^2},\\
 c_2(\sigma,n)=&\frac{10 (\sigma +n  )^2- 1}{8  (\sigma +n  )^4 (2 \sigma +2 n  -1 )^2 (2 \sigma +2 n  +1 )^2}.  \ea\ee
Therefore in the four-dimensional limit, the original shift in the chemical potential \eqref{sd2} translates into a shift of the parameter $\sigma$ by 
\be \sigma \to \sigma +n.\ee
It follows that
\be \label{lambdain}  \exp [\mathsf J^{\rm inst}(q^{2n}Q_f, q^{2n}Q, \hbar)]\quad \xrightarrow{4D} \quad B(T, \sigma+n), \ee
where \be B(T, \sigma)=\left( 1+\frac{T}{2 \sigma ^2} +\frac{\left(8 \sigma ^2+1\right) T^2}{4 \sigma ^2 \left(1-4 \sigma ^2\right)^2}+\mathcal{O}(T^3)\right).\ee
This is the well known  instanton contribution to the Nekrasov partition function as computed in \cite{n,Bruzzo:2002xf,Flume:2002az} and it defines a convergent series in $T$.
As mentioned earlier, in \cite{ghm} $m$ and  $\hbar$  are positive numbers, therefore 
$ \sigma$ is purely imaginary.  It follows that  $B(T, \sigma)$ has no poles in the domain of interest.

\subsubsection{One-loop contribution }
The four-dimensional limit of the one-loop contribution can be computed straightforwardly  from the integral representation \eqref{intol}. We have
\be \label{4dol}  \ba \exp [\mathsf J^{\rm oneloop}(q^{2n}Q_f, \hbar)]\quad \xrightarrow{4D}& \quad \frac{2 \zeta (3)}{\beta ^2\epsilon^2}-\frac{2 \left(\pi ^2 (\sigma +n) \epsilon \right)}{3 \beta \epsilon^2} +6  (\sigma +n) ^2-4  (\sigma +n) ^2 \log (2 \beta   (\sigma +n) \epsilon) 
\\
&+ \frac{\log (\beta )}{6}+{\rm Re} \int_0^{\re^{\ri 0}\infty}\rd x \frac{ x}{e^{2 \pi  x}-1}\log \left({4   (\sigma +n) ^2 \epsilon^2 +\epsilon^2  x^2}\right)^2.\ea\ee
Moreover \footnote{ The contour integral is reminiscent of a lateral Borel resummation see \cite{ho2,hm} for more details. We have checked it numerically. } 
\be \label{toshow}\ba  -\log (G(2 \sigma +1) G(1-2 \sigma )) &= {\rm Re} \int_0^{\re^{\ri 0}\infty}\rd x \frac{ x}{e^{2 \pi  x}-1} \log \left({4  \sigma ^2  +  x^2}\right)^2-2 \zeta'(-1)\\
&+6 \sigma ^2-\sigma ^2 \log \left(2 \sigma \right)^4,\ea\ee
where $G(z)$ is the Barnes G--function.
Notice that in our convention $\sigma$ is purely imaginary. 
We get
\be \label{4d1} \exp [\mathsf J^{\rm oneloop}(q^{2n}Q_f, \hbar)]\quad \xrightarrow{4D} \quad  \re^{ \frac{2 \zeta (3)}{\beta ^2 \epsilon ^2}+ 2\zeta'(-1)}{(\beta^4 \epsilon^4)^{-(\sigma+n)^2+1/24}\over  G(1-2(\sigma+n))G(1+2(\sigma+n))}\re^{-\frac{2 \pi ^2 (\sigma+n) }{3 \beta  \epsilon }}.\ee

\subsubsection{Constant map contribution }
By using the explicit expression \eqref{cm} it is easy to see that
\be\label{cm4d} \ba  \re^{A(m,\hbar)} \quad \xrightarrow{4D} \quad & (\beta^4 \epsilon^4 T)^{\sigma^2} \re^{4 \sqrt{T}}{\re^{-2 {\zeta(3)\over \beta^2 \epsilon^2}+\frac{2 \pi ^2 \sigma }{3 \beta  \epsilon }}}\re^{\zeta'(-1)}\re^{\frac{1}{12} \log (2) }T^{-1/48}\\
 \times  & \left(\beta ^4 T \epsilon ^4\right)^{\frac{\log ^2\left(\beta ^4 T \epsilon ^4\right)+4 \pi ^2}{12 \beta ^2 \epsilon ^2}} e^{\frac{\sigma  \log ^2\left(\beta ^4 T \epsilon ^4\right)}{2 \beta  \epsilon }} \ea \ee
Similarly 
\be \label{4dp}\ba \re^{\mathsf P(t(\mu-2\pi \ri n, \hbar), m, \hbar)}  \quad \xrightarrow{4D} \quad &  \left(T\beta^4 \epsilon^4\right)^{-1/24} \exp\left[n\frac{2 \pi ^2}{3 \beta  \epsilon }\right]\left(\beta ^4 T \epsilon ^4\right)^{2 \sigma n +n^2} \\
&\times \left(\beta ^4 T \epsilon ^4\right)^{-\frac{\log ^2\left(\beta ^4 T \epsilon ^4\right)+4 \pi ^2}{12 \beta ^2 \epsilon ^2}} e^{-\frac{\sigma  \log ^2\left(\beta ^4 T \epsilon ^4\right)}{2 \beta  \epsilon }}.\ea \ee

\subsubsection{The  four--dimensional limit of topological string }
By combining together \eqref{cm4d} , \eqref{4dp}, \eqref{4d1}, \eqref{lambdain}
 we get
\be \label{sd4d} \ba \Xi^{\rm 4d}_{\rm S}(\sigma,T)=&\re^{\frac{\log (2)}{12} +3\zeta'(-1)}T^{-1/16}\re^{4 \sqrt{T}} \sum_{n \in \IZ}  {T^{(\sigma+n)^2} B(T, \sigma+n) \over  G(1-2(\sigma+n))G(1+2(\sigma+n))}.\ea\ee
This is the four-dimensional limit \eqref{ftl} of the conjectural expression \eqref{sd2} for the spectral determinant \eqref{sd1}.  We add the subscript $S$ in \eqref{sd4d} to stress that this is the spectral determinant computed from the string theory side of the conjecture.

It was first conjectured and then proved \cite{ilt,gil,bes,ilte}  that 
\be \label{GIL}\widehat \tau(T, \sigma, \eta)= \sum_{n \in \IZ}  \re^{4 \pi \ri n \eta}{T^{(\sigma+n)^2} B(T, \sigma+n) \over  G(1-2(\sigma+n))G(1+2(\sigma+n))}\ee
satisfies  the  Painlev\'e $\rm{III}_3$ equation in the $\tau$ form. %
   This means that
\be \re^{-\ri \widehat U(t, \sigma, \eta) }={4 \over t }{d \over dt} t {d \over dt}\log\widehat \tau(t^4 2^{-12},\sigma, \eta)\ee
satisfies 
\be\left(\left({d\over dt}\right)^2+{1\over t}{d\over dt}\right)   \widehat U(t, \sigma, \eta) =-\sin \left[  \widehat U(t, \sigma, \eta)\right]. \ee
Notice that \eqref{GIL} is precisely the dual Nekrasov--Okounkov  partition function \cite{no2}.
The variables $(\sigma, \eta)$ are the monodromy data of the related Fuchsian system, which fix the initial conditions of $\widehat U$. 
They can be read off from the asymptotic expansion at small $t$ \cite{jimbo,nov}
\be \label{eq1}   e^{\ri \widehat U(t, \sigma, \eta) }\sim-\re^{4\pi \ri \eta}\frac{\Gamma^2\left(1-2\sigma\right)}{\Gamma^2\left(2\sigma\right)}\left(\frac{t}{8}\right)^{8\sigma-2}.
\ee
The solution \eqref{GIL} is a convergent, well-defined function whenever $2\sigma \notin \IZ$ \cite{ilt,bes}. Due to the periodicity properties it is enough to consider $\sigma =0, 1/2$, for these values there exists a regularization procedure leading to a well-defined solution with $\eta \to 0$
 \footnote{ We thank Oleg Lisovyy for a discussion on this point.}.

Using these results, we conclude that  the conjectural expression \eqref{sd2} in the four-dimensional limit reproduces the $\tau$ function of Painlev\'e $\rm{III}_3$, namely
\be\label{tauxi}  \tau(T,\sigma)= \re^{-4 \sqrt{T}}  \Xi^{\rm 4d}_S(\sigma,T)=\re^{\frac{\log (2)}{12} +3\zeta'(-1)}T^{-1/16}\widehat \tau(T, {\sigma}, 0),\ee
with initial conditions \footnote{We use the notation of \cite{ilt}.}
\be (\sigma, \eta=0). \ee
To  completely fix  the $\tau$ function we also have to specify the small $T$ expansion. We have
\be\label{sT} \ba \re^{-4\sqrt{T}}  \Xi_S^{\rm 4d}(\sigma,T)\approx  T^{\sigma ^2-\frac{1}{16}} \re^{{1\over 12} \log 2+3 \zeta'(-1)}{1\over G(1-2\sigma)G(1+2\sigma)} .
\ea\ee
Notice that in our set up $\ri \sigma \in \IR/\{0\}$, hence we do not have any problems with discontinuities or poles.

\subsection{The operator theory computation}\label{4dope}
In this section we  focus on the operator side of the conjecture \cite{ghm}.  We would like to show that the spectral determinant \eqref{sd1} for the operator
\be \label{rhop1p1b}\rho_{\IP^1 \times \IP^1}=\left( \re^{\hat x}+\re^{\hat p}+\re^{-\hat p}+m_{\mathbb{F}_0}\re^{-\hat x}\right)^{-1}, \quad [\hat x,\hat p]=\ri \hbar \ee
 is related to the $\tau$ function of Painlev\'e $\rm{III}_3$  as in \eqref{tauxi}. We  do this without using the conjectural expression \eqref{sd2}. The strategy relies on the work of \cite{zamo} which is summarized in appendix \ref{zamosd}.

The kernel of the operator \eqref{rhop1p1b} has been computed in \cite{kmz} and it reads
 \be\label{kp1p1} \rho_{\IP^1\times \IP^1} (y_1,y_2)= \re^{-\pi {b} \gamma /2}  {|f(y_1)| |f(y_2)| \over  2{b} \cosh\left(\pi {y_1- y_2 \over {b}}  \right) },
\ee
where
\be f(x) =\re^{\pi x {b}/2}{\fad(x-\gamma/2+ \ri b/4)\over \fad(x+\gamma/2- \ri b /4)}, \qquad \hbar=\pi b^2. \ee
We denote by $ \fad$ the Faddeev's quantum dilogarithm \cite{faddeev,fk}. 
The parameter $\gamma$ in \eqref{kp1p1} is related to the mass $m$ in \eqref{smallm}, more precisely we have
\be \gamma={b\over 4 \pi} \log m.\ee
The small $\kappa$ expansion of the spectral determinant describing the operator \eqref{rhop1p1b}  is given by
\be \label{sdp1p1} \Xi_{\IP^1\times \IP^1}(\kappa, \hbar)=\sum_{N\geq 0} \kappa^N{Z(N, \hbar)}, \quad \kappa=\re^{\mu},\ee
where $Z(N, \hbar)$ can be written as a matrix model  of the form \cite{kmz}
\be 
\label{mmo}Z(N,\hbar)={\re^{- \frac{ \hbar}{8 \pi } N \log m}\over N!}\int \frac{\rd^N z}{(2\pi)^N} \re^{- \sum_{i=1}^N (V(z_i,\hbar))} \frac{\prod_{i < j} (z_i-z_j)^2}{\prod_{i , j} (z_i+z_j)}.
\ee 
The potential $V(z, \hbar)$  is given by 
\be \ba & V(z, \hbar)=- \log\left |f\left (\frac{b \log {z}}{2\pi} \right) \right|^2.
\ea\ee
We are interested in computing the four-dimensional limit of
\be \mathcal{V}^{\rm tot}(\hbar,m, \mu)=-{\hbar \over 8 \pi}\log m+\left(\mu-{t(\mu, \hbar)  \over 2}\right)+{t(\mu, \hbar)  \over 2}+\log\left | f\left (\frac{b \log {z}}{2\pi}  \right) \right|^2.\ee
It is useful to write the quantum dilogarithm in terms of double sine function $s_b$ as in \cite{hel}. 
We have
\be\ba  \left |  f\left (\frac{b \log {z}}{2\pi}  \right) \right|^2=&\left | \re^{ \frac{b^2 \log {z}}{4}  }{\fad(\frac{b \log {z m^{-1/4}}}{2\pi} + \ri b/4)\over \fad(\frac{b \log {z m^{1/4}}}{2\pi}- \ri b /4)} \right|^2=\left | {s_b(\frac{b \log {z m^{-1/4}}}{2\pi} + \ri b/4)\over s_b(\frac{b \log {z m^{1/4}}}{2\pi}- \ri b /4))} \right|^2 \\
=& {s_b(\frac{b \log {z m^{-1/4}}}{2\pi} + \ri b/4)\over s_b(\frac{b \log {z m^{-1/4}}}{2\pi} - \ri b/4)}  { s_b(\frac{b \log {z m^{1/4}}}{2\pi}+ \ri b /4)\over s_b(\frac{b \log {z m^{1/4}}}{2\pi}- \ri b /4))}, \ea \ee
where we used
\be \overline{s_b(z)}={1 \over {s_b(\overline z)} }.\ee
The integral form of the double sine is  \cite{ho2} 
\be  \label{dsf} \ri \log s_b(z)=\frac{\pi z^2}{2}+\frac{\text{Li}_2(-\re^{2 \pi bz})}{2\pi b^2}
+\int_0^\infty \frac{\rd x}{\re^{2\pi x}+1}
\log\left(\frac{1+\re^{b2 \pi z-2\pi b^2x}}{1+\re^{b2 \pi z+2\pi b^2x}}\right).\ee
Using \eqref{dsf} together with
\be s_b(x)=s_{b^{-1}}(x) \ee
we obtain 
\be  \re^{\mathcal{V}^{\rm tot}(\hbar,m, \mu)- \left(\mu-{t(\mu, \hbar)  \over 2}\right)} \quad \xrightarrow{4D} \quad \re^{-4 T^{1/4} \left( z+{1\over z}\right)}\re^{-2 \pi \ri \sigma}.\ee
The results of appendix \ref{ap2} together with the dictionary \eqref{ftl} lead to
\be  \mu-{t(\mu, \hbar)  \over 2} \quad \xrightarrow{4D} \quad \log \left( 1+ \re^{4\pi \ri \sigma}\right).\ee
Hence we obtain 
\be \re^{\mathcal{V}^{\rm tot}(\hbar,m, \mu) }\quad \xrightarrow{4D} \quad \re^{-4 T^{1/4} \left( z+{1\over z}\right)}\re^{\log\left( 2 \cos(2\pi \sigma)\right)}.
\ee
Therefore in the four-dimensional limit we can write the spectral determinant \eqref{sdp1p1} as
\be \label{op4d} \Xi^{\rm 4d}_O(\sigma, T)= \sum_{N\geq 0} \left(  {\cos 2 \pi \sigma\over 2 \pi}\right)^N Z_{\rm 4d}(N,T), \ee
where
 \be \label{Z4d} Z_{\rm 4d}(N,T)= {1 \over N!}\int \prod_{i=1}^N  \frac{\rd z_i}{z_i} \re^{- 2 \sum_{i=1}^N ( V_{\rm 4d} (z_i,T))} \frac{\prod_{i < j} (z_i-z_j)^2}{\prod_{i < j} (z_i+z_j)^2},\ee
 \be \label{pt}  V_{\rm 4d} (z_i,T)=2 T^{1/4}\left(z_i +{1\over z_i}\right).\ee
 We add the subscript $O$ in \eqref{op4d} to stress that this is the spectral determinant computed in the operator theory side of the conjecture. 
This is precisely a spectral determinant of the Zamolodchikov form  \eqref{zamomm}. It follows that 
the four-dimensional limit of \eqref{sd1}, for the operator \eqref{rhop1p1b}, computes  the $\tau$ function of Painlev\'e $\rm{III}_3$ according to
\be \label{tauop}\tau(T,\sigma) =\Xi_O^{4d}( \sigma,T)\re^{-4 \sqrt{T} }.\ee
It is well known that the  Barnes G--function is related to the polygamma function of negative order $\psi^{(-2)}$  as \be\label{ident2} \log G(z)=\frac{1}{2} z \log (2 \pi )-\psi ^{(-2)}(z)+(z-1) \left(\log \Gamma (z)-\frac{z}{2}\right), \quad {\rm Re}(z)>0.\ee
By using \eqref{zamoBG} together with \eqref{ident2}
it is straightforward to see that the small $T$ asymptotic of \eqref{tauop} is given precisely by \eqref{sT}.

Summarizing, we have shown that, in the four-dimensional limit, both the spectral determinant \eqref{sd1} and its conjectural expression \eqref{sd2}  satisfy the Painlev\'e $\rm{III}_3$ equation in the $\tau$ form with the same monodromy conditions given by
\be (\sigma, \eta=0), \ee
and the same asymptotic expansion \eqref{sT}. Therefore, within the dictionary \eqref{ftl} and in particular for $\sigma \in \ri \IR/\{0\}$, and $T>0$,   we have   \be\label{fin} \Xi^{\rm 4d}_O( \sigma,T)= \Xi^{\rm 4d}_S(\sigma,T),\ee
which 
concludes our proof.

\section{Gauge theory and quantum gas} \label{gauge}
In this section we focus on the gauge theory interpretation of the above results.
\subsection{The spectral problem in four dimensions}\label{spectral4}
  The  computation  of Section \ref{sp} shows that the spectrum of the operator \eqref{opri} is determined by the  zeros of the tau function \eqref{GIL}. This in turn determines the spectrum of an ideal Fermi gas.

Let us write the four-dimensional matrix model \eqref{Z4d} as
 \be \label{coshmm}  Z_{\rm 4d}(N,T)= {1\over N!} \int \prod_{i=1}^N  {\rd x_i} \re^{-8 T^{1/4} \cosh(x_i)}\prod_{i < j}\tanh^2({x_i-x_j \over 2}).\ee
This is the so--called polymer matrix model studied in \cite{zamo,gm,fadsal,cfiv} (see chapter 20 of \cite{mussardo} for a survey on the subject).
 It was found in \cite{gm} that this matrix model describes an ideal Fermi gas whose density matrix is
 \be \label{rhopoly}\rho(x_1, x_2)= {\re^{-{4 T^{1/4} }\cosh(x_1) - {4 T^{1/4} }\cosh(x_2)} \over 
 \cosh\left( {x_1 - x_2\over 2 } \right)}. 
 \ee
Similarly to what we did in Section \eqref{4dope} we find that, in the four-dimensional limit, the kernel \eqref{kp1p1} reproduces the kernel \eqref{rhopoly}. 
 The  operator corresponding to \eqref{rhopoly} is
 \be \label{opr} \rho= \re^{-{4 T^{1/4} } \cosh(\hat x)}{4 \pi \over \left(\re^{\hat p/2}+ \re^{-\hat p/2} \right)}\re^{-{4 T^{1/4} } \cosh(\hat x)}, \qquad [\hat x, \hat p]= 2 \pi \ri .\ee
As discussed in \cite{gm} this corresponds to an ideal Fermi gas in an external  potential
\be V(x)= {8 T^{1/4}}\cosh(x) \ee
and with a non--standard kinetic term given by 
\be T(p)=\log \left[2 \cosh({p/2})\right].\ee
 To get some physical intuition into this system it is useful to consider the large energy limit of this gas, i.e.~$p,x$  large. In this regime the Hamiltonian  can be approximated by
 \be H(p, x) \approx {1\over 2}\mid p\mid+ {4 T^{1/4}} \re^{\mid x\mid}.\ee 
 Therefore we have an ultra-relativistic Fermi gas in a confining potential.
 
 The operator \eqref{opr} is of trace class and has a positive, discrete spectrum $\re^{-E_n}$ which is determined by  the vanishing locus of
 \be \label{fermiok}\Xi_{\rm O}^{\rm 4d}(\kappa,T)= \prod_{n\geq 0}\left( 1+\kappa \re^{-E_n}\right)=\sum_{N\geq 0}\kappa^N  Z_{\rm 4d}(N,T).\ee
 In statistical mechanics one typically refers to $\Xi_{\rm O}^{\rm 4d}(\kappa,T)$ as the  {grand canonical partition function} of the gas. 
To make contact with the Painlev\'e $\rm{III}_3$ dictionary we have
 \be 2 \pi \kappa =  \cos (2 \pi \sigma). \ee
 Therefore the region which is interesting from the spectral theory point of view is parametrized by \footnote{
 In Section \ref{sp} we restrict ourself to $\ri \sigma \in \IR/\{0\} $ to make contact with the topological string parameters. However from the four-dimensional perspective we can take more general values of $\sigma$ as in \cite{zamo} and in \cite{ilt1,ilt,gil1,gil,bes,ilte}.   From the topological string perspective this translates into the necessity of extending the conjecture \cite{ghm} to arbitrary complex values of $\hbar, m$. In particular notice that for  $2\sigma \in \IZ+\ri \IR/\{0\} $ one can still invert the mirror map as in appendix \ref{ap2} and \eqref{wan} still vanishes.
 }
 \be \sigma={1\over 2}+\ri \sigma_r, \qquad \sigma_r \in \IR/\{0\}.\ee
Using this dictionary, we can compute numerically the zeros of \eqref{sd4d} which give  the spectrum of the Fermi gas described by \eqref{rhopoly}. More precisely we have
\be \label{spe}\left\{{E_n}\right\}_{n=0,1,\dots}= \left\{ \log  \left[{1\over 2\pi}  \cosh (2 \pi \sigma_r^{(n)})\right]  :    \Xi^{\rm 4d}_S({1\over 2}+\ri \sigma_r^{(n)},T)=0 \right\}.\ee
In Table \ref{tb1} we compare the numerical spectrum  of the operator \eqref{rhopoly} with the zeros of  \eqref{sd4d}: we find perfect agreement.  Therefore  the vanishing locus of the $\tau$ function which solves the Painlev\'e $\rm{III}_3$ equation gives the spectrum of the operator \eqref{rhopoly}, as we showed analytically in Section \ref{sp}.
%
%
 \begin{table}[t] 
\centering
   \begin{tabular}{l l l}
  \\
Order& $E_0 $  & $E_1$\\
\hline
 1 & \underline {0.56899}29450193& \underline {2.7765099}480066\\
 2 & \underline {0.56899302}27978& \underline {2.776509963}4917\\
 4 & \underline {0.5689930268761} & \underline {2.7765099636767}\\ 
 \hline
Numerical value &     $0.5689930268761$  & $ 2.7765099636767 $ \\
\end{tabular}
\\
\caption{ The  first two energy levels for the operator \eqref{rhopoly}, 
 obtained from the vanishing locus of the $\tau$ function \eqref{sd4d} as explained in equation \eqref{spe}. The expression of $\Xi^{\rm 4d}_S(\sigma,T)$ is given as a convergent 
series  at small $T$. As we keep more terms in the series expansion we quickly approach the energy obtained 
by using the numerical methods of \cite{ghmabjm} applied to the operator \eqref{rhopoly}. We take $T^{1/4}=\frac{ \pi }{21}$.
}
 \label{tb1}
\end{table}

If we consider the gauge theory perspective, the results of Section \ref{sp} show that the spectrum of \eqref{rhopoly} is computed by the four-dimensional Nekrasov partition function \cite{n}
\be  Z^{\rm Nek}(\epsilon, a,\Lambda )\ee
where the equivariant parameters are set to $\epsilon_1=-\epsilon_2= \epsilon$.
 Indeed we have
\be Z^{\rm Nek}(\epsilon, a,\Lambda )=Z(\sigma,T)= {T^{\sigma^2} B(T, \sigma) \over  G(1-2\sigma)G(1+2\sigma)}, \ee
where \be \sigma= {a / \epsilon}, \qquad T={\Lambda^4 \over 2^4 \epsilon^4 }. \ee
In \cite{no2} the authors  introduced the dual partition function as
\be \label{NO2} Z^{\rm NO}(\eta,\sigma,T)= \sum_{n\in \IZ} \re^{4 \pi \ri n \eta}  Z(\sigma+n,T). \ee
Our analysis shows that this dual partition function, at $\eta=0$, corresponds to the grand canonical partition function of an ideal Fermi gas
whose density matrix is given by  \eqref{rhopoly}. In this correspondence 
the Seiberg--Witten period $\sigma$  and the instanton counting parameter $T$  correspond to the chemical potential of the gas and the strength of the external potential respectively. 

\subsection{A matrix model for Nekrasov's partition function } \label{gauge2}
In the above subsection we  focused on the grand canonical ensemble  in  which the grand canonical partition function $\Xi^{\rm 4d}$ of the gas makes contact with the dual NO partition function. In this section, instead, we study the canonical ensemble, in particular we show that the canonical partition function $Z^{\rm 4d}$ can be identified with the Nekrasov partition function in the magnetic frame.  

It is well known \cite{n,no2,sw1,sw2}  that the  4-dimensional Nekrasov partition function reproduces the Seiberg-Witten prepotential $\mathcal{F}_0(a, \Lambda)$ and its gravitational corrections  \footnote{ These can also be computed from the five-dimensional perspective see for instance \cite{dgkv,cv,dv3,kmt,ikp}.} $\mathcal{F}_g(a, \Lambda)$  in the expansion
\be \label{zkp}Z^{\rm Nek}(\epsilon, a,\Lambda )= \exp \left[ {\sum_{g\geq 0} \epsilon^{2g-2}\mathcal{F}_g(a, \Lambda)}\right],\ee 
where $\epsilon$ is the vacuum expectation value of the self-dual graviphoton field strength. In this section we show that the free energies $\mathcal{F}_g$ emerge when we study the so--call 't Hooft expansion of \eqref{Z4d}.

The  't Hooft regime of \eqref{Z4d} is defined as \cite{gm} 
 \be \label{thoofta} N,T \to \infty, \qquad {N\over 2 T^{1/4} } =  \lambda \quad {\rm fixed}, \ee
and  \be \label{gex}\log Z_{\rm 4d}(N,T)= \sum_{g \geq 0}(2T^{1/4})^{2-2g}F_g^{\rm 4d}(\lambda).\ee
In general it is difficult to compute  the genus expansion \eqref{gex} directly in a matrix model, i.e. without using tools such as the holomorphic anomaly.
However, for the $O(2)$ matrix models,  exact formulae for $F_0$ and $F_1$ exist in the literature \cite{ek1,ek2,kos40,gm}. For the particular potential \eqref{pt} one can show that  \eqref{Z4d} is a one cut matrix model with endpoints $(a,a^{-1})$ where $a$ is related to $\lambda$ 
as \cite{gm}
\be \label{Tcoshex} 
\lambda (a)=\frac{-\pi +\left(2 E(k)+\left(-1+k^2\right) \kk\right) \kkp}{ k^{1/2} \pi  \kk} \qquad  k=a^2,\ee
and we denote by $K,E$ the elliptic integral of first and second kinds.
For small values of $\lambda$ one has
\be\label{als} a=1-\sqrt{\lambda }+\frac{\lambda }{2}-\frac{\lambda ^2}{16}-\frac{\lambda ^{3/2}}{16}+\mathcal{O}(\lambda^{5/2}).\ee
Moreover 
\be {\rd^2 F_0^{\rm 4d} (\lambda)\over \rd \lambda^2}=-2\pi {K(k) \over  K'(k)},\ee
 which leads to
 \be\label{F0small} F_0^{\rm 4d} (\lambda)=\frac{1}{2} \lambda^2 \left( \log\left({\lambda\over 16}\right) -\frac{3 }{2} \right)-4\lambda -\frac{\lambda^3}{16}+\frac{5 \lambda^4}{512}-\frac{11 \lambda^5}{4096}+\mathcal{O}(\lambda^6).\ee
This is precisely  the genus zero free energy of Seiberg--Witten theory $\mathcal{F}_0$ in the magnetic frame  \cite{hk06} where we identify the 't Hooft parameter $\lambda$ with the dual period \footnote{See \cite{hk06} for the explicit definition of $\tilde a_D$.} $\tilde a_D$ of the Seiberg--Witten curve \be \label{lasad} \lambda = -\tilde a_D.\ee
Moreover, as in \cite{hk06}, we set the Seiberg--Witten scale to be $\Lambda=1$.  Therefore in this correspondence the parameters describing the canonical ensemble of the gas, namely the  number  of particles $N$ and the strength of the potential  $T$, get mapped to the dual period $\tilde a_D$ and  the graviphoton strength  $\epsilon$. 
Similarly by specializing the general formula for $F_1$ of  \cite{ek1,gm}  to the potential \eqref{pt} we find
\be F_1^{\rm 4d}(\lambda)=-\frac{1}{4} \log \left(K\left(1-\frac{1}{a(\lambda)^4}\right) K\left(1-a(\lambda)^4\right)\right)-\frac{1}{6} \log \left(\frac{1}{a(\lambda)^2}-a(\lambda)^2\right)+{\rm constant}, \ee
where the constant is fixed by the Gaussian behavior of the matrix model \eqref{Z4d} at small $\lambda$.
By using \eqref{als} we obtain
\be F_1^{\rm 4d}(\lambda)= -\frac{1}{12} \log \left(\frac{\lambda }{4}\right)+\zeta'(-1) +{\lambda \over 32} - {3 \lambda^2\over 512 }+ {19 \lambda^3\over12288} - {
 3 \lambda^4 \over 8192}+ \mathcal{O}(\lambda^5).\ee
 This reproduces the genus one free energy of Seiberg--Witten theory $\mathcal{F}_1$ in the magnetic frame   \cite{hk06} with the identification \eqref{lasad}. Even though we do not have a direct all order proof, we expect this identification to hold also for higher $F_g^{\rm 4d}$'s.  Therefore in our formalism the perturbative expansion \eqref{zkp} of the Nekrasov partition function in the magnetic frame emerges in the 't Hooft limit \eqref{gex} of the gas. 
 In particular this implies  we can now solve the $O(2)$ matrix model \eqref{Z4d} at all order in the genus expansion by using the holomorphic anomaly equations of the gauge theory.

Notice that the matrix model \eqref{Z4d} provides an expression for the spectral determinant    \eqref{sd4d}
which resums the expansion  in $T$. Moreover the 't Hooft expansion of the spectral determinant  \eqref{op4d} naturally reproduces the long distance solution to Painlev\'e ${\rm III}_3$ proposed in \cite{ilt}. A detailed discussion of this regime will appear in \cite{blmst}.

 Other proposals for a matrix model description of the Nekrasov partition function appeared before in the literature as for instance \cite{Dijkgraaf:2009pc,Eguchi:2009gf,Eguchi:2010rf, Bonelli:2011na,Bonelli:2010gk,Maruyoshi:2014eja}. These are constructed by expressing the conformal blocks  as matrix models and are different from our which arises from the quantization of mirror curves to toric CYs. In particular the perturbative expansion of \cite{Bonelli:2011na} reproduces the  $SU(2)$ gauge theory in the NS background ($\epsilon_2=0$) while in our model we recover the perturbative expansion in the self--dual background.
  Similarly in \cite{Dijkgraaf:2009pc,Eguchi:2009gf,Eguchi:2010rf} a Penner-like matrix model was proposed to describe $SU(2)$ gauge theory with  hypermultiplets in a self--dual background. It would be interesting to extend our construction to the case of  gauge theory with   hypermultiplets and compare with these proposals more in detail.
  
  From a pure mathematical point of view, the main difference between  previous proposals and the one we are considering here consists in  the interaction term. In our model the interaction is the one of an $O(2)$ model  while in previous proposals it was given by the Vandermonde determinant\footnote{ or its $\beta$ deformed generalization.}. Moreover the standard 't Hooft expansion of \eqref{Z4d} describes Seiberg--Witten theory in the magnetic frame and not the electric one as it was the case in previous proposals. 
In turns, these differences are related to the fact that our  model arises as a four-dimensional limit of the matrix model describing topological string on toric CYs \cite{ghm, kmz}. Other proposals instead are more related to topological string theory on the Dijkgraaf--Vafa types of manifold \cite{dv} which can also be used to engineer  Seiberg--Witten theory in four dimension \cite{kkv,ikp,kmt,dv3,dgkv,cv}.

 \section{Conclusion}\raggedbottom

In this paper we proposed  a  new double-scaling limit in which the spectral determinant of \cite{ghm} makes contact with the dual  four-dimensional $SU(2)$ 
 Nekrasov--Okounkov partition function and with the Painlev\'e $\rm{III}_3$ equation.
We used the recent developments in the context of Painlev\'e $\rm{III}_3$  \cite{ilt1,ilt,gil1,gil,bes,ilte}, together with 
previous works on spectral determinants \cite{zamo}, to prove the conjecture of \cite{ghm} in 
this new double-scaling limit of local $\IP^1 \times \IP^1$. 
Within this construction, we  
showed that the vanishing locus of the  $\tau$ function computes the  spectrum of an ideal Fermi gas.
  
In addition, the partition function of this gas can be written as an $O(2)$ matrix model which provides an exact expression for the Nekrasov partition function in the magnetic frame and in the self--dual $\Omega$--background. 
Interestingly, this gas
 appears as well in the  computation of two-point functions for the 2d Ising model  and in the study of 2d polymers \cite{mussardo,zamo, ceva,cfiv,fadsal}. 
 As a consequence, one can compute Nekrasov's partition function exactly by using the TBA  techniques of \cite{fadsal,cfiv,zamo}. 
 It would be interesting to explore
this  relation in more detail in view of the results presented here. 
We hope to report on this in the near future.

Furthermore, it was conjectured in \cite{ns} that in the NS background ($\epsilon_2=0$) the gauge theory is strictly related to quantum integrable models. Our results show 
that there is an underlying quantum mechanical system also in the self--dual $\Omega$--background. Moreover, we proved that the spectral determinant, and therefore the spectrum, of this system is computed by the NO partition function.

In this work, we focused on  the pure  $SU(2)$  $\mathcal{N}=2$ four-dimensional gauge theory. However,  it would be interesting to combine the techniques developed here with the work of \cite{blmst} to study  the four-dimensional gauge theory  with matter. In particular, our approach can be used to construct explicit Fredholm determinant solutions to other Painlev\'e equations. To do this, the only task is to compute the kernel for the operators associated with the mirror curves that engineer  gauge theories  with matter by following the prescription in  \cite{ghm,kama}. The four-dimensional limit \eqref{ftl} of these operators should provide the desired Fredholm determinant  solution.

Most importantly, the gauge theory/Fermi gas correspondence presented here can be understood from a purely four-dimensional perspective by using lattice quantization. 
Indeed, the operator \eqref{opri} can be constructed directly  from the Seiberg--Witten curve by using an unusual quantization scheme: the so--called  lattice quantization scheme \cite{parisi}.
This opens the road for an immediate generalization of the Fermi gas formalism of \cite{mp} to other gauge theories with matter multiplets and it allows to construct explicit Fredholm determinant  solutions to others Painlev\'e equations. A detailed discussion of this generalization will be presented elsewhere.  

Moreover, the case of higher rank gauge theory should also be investigated. In this context, one can deform the theory 
to the generating function of local BPS observables $\tr \Phi^k$, where $\Phi$ is the scalar component of the $\mathcal{N}=2$ vector multiplet.
It would be interesting to understand the counterpart of these deformations from the spectral theory point of view.

Notice that the general solution to Painlev\'e $\rm{III}_3$ equation contains an additional parameter $\eta$ which is set to zero in our construction. This suggests that a generalization of the spectral determinant may exist, including this extra parameter. It would be interesting to investigate its role in the context of topological string.

Finally, a natural extension of this work is the generalization to the  five-dimensional case describing the full topological string amplitude. 
We will report on this in the near future \cite{progress}.

\section*{Acknowledgements}
We would like to thank Davide Guzzetti, Yasuyuki Hatsuda, Oleg Lisovyy, Marcos Mari\~no, Massimiliano Ronzani  and Antonio Sciarappa for useful discussions  and for clarifications on their previous works.
Especially Omar Foda, Marcos Mari\~no and Antonio Sciarappa for useful comments and a careful reading of the manuscript.
This research was partly supported by the INFN Research Projects GAST and ST$\&$FI and by PRIN "Geometria delle variet\`a algebriche".
   
\appendix

\section{Quantum A-period}\label{ap2}
The notion of quantum A-period  was studied in the context of AGT correspondence \cite{mirmor,mirmor2,mt,Bonelli:2011na,Tan:2013tq} and topological strings 
 \cite{adkmv,acdkv}. It is the integral of a quantum differential  over the A-cycle of a given curve. In the particular case of  local $\IP^1 \times \IP^1$ the quantum A-period has  been computed in \cite{acdkv}  and reads
\be{t(\mu, \hbar) \over 2}= \Pi_A(\mu, \hbar )=\mu +(-m_{\IF_0}-1) z+z^2 \left(-\frac{3 m_{\IF_0}^2}{2}-m_{\IF_0} q-\frac{m_{\IF_0}}{q}-4
   m_{\IF_0}-\frac{3}{2}\right)+O\left(z^3\right),\ee
   where $z=\re^{-2\mu}$ and $q=\re^{ \ri \hbar}$.
   This relation can also be inverted by using an ansatz of type
   \be \mu=\Pi_A+\sum_{n\geq 1} a_n(m_{\IF_0})\re^{-2n\Pi_A}.\ee
   We find
   \be  \mu=\Pi_A+\sum_{n\ge 0} \Pi_n(z_2,q)z_1^n\ee
   where
  \be z_1=\re^{- 2\Pi_A}, \qquad  z_2=m_{{\IF_0}}\re^{- 2\Pi_A}. \ee
  Notice that in the 4d limit \eqref{ftl} we have \be z_1 \to 0, \qquad z_2 \to \re^{4 \pi \ri \sigma}.\ee
  Therefore it is important to resum the $z_2$ expansion. For the first few coefficients we find
  \be  \ba  \Pi_0(z_2,q)=&\log (1+z_2) ,
  \\
    \Pi_1(z_2,q)=&1+\frac{z_2 \left(-q^2+2 q z_2-1\right)}{(z_2-1) (z_2-q) (q z_2-1)}.\ea\ee 
    In the four-dimensional limit we have
      \be  \ba  \Pi_0(z_2,q)&\quad \xrightarrow{4D}\quad   \log (1+\re^{4 \pi \ri \sigma}),
  \\ \Pi_1(z_2,q)& \quad \xrightarrow{4D}\quad 1+ \frac{1}{\cos (4 \pi  \sigma )-1}.\ea\ee 
  Notice that in our construction the mass parameter of local $\IP^1 \times \IP^1$ and $\hbar$ are both positive. Hence  $\sigma$ is purely imaginary and $\sigma\neq0$, therefore $ \Pi_1(z_2,q)$ is perfectly well-defined. Similarly for the other $ \Pi_n(z_2,q)$ .
It follows that
  \be  \mu- {t(\mu, \hbar) \over 2}\quad \xrightarrow{4D}\quad   \log (1+\re^{4 \pi \ri \sigma}) . \ee

\section{Standard and NS free energies}\label{ap1}
The free energy of standard topological string  at large radius was computed in \cite{gv,akmv2} and it reads
\be \label{fgv}F^{\rm GV}({\bf{t}}, g_s)=\sum_{g\ge 0} \sum_{\bf d} \sum_{w=1}^\infty {1\over w} n_g^{ {\bf d}} \left(2 \sin { w g_s \over 2} \right)^{2g-2} \re^{-w {\bf d} \cdot {\bf t}}  .\ee
The variable ${\bf{t}}$ denotes the K\"ahler parameters of the geometry, $g_s$ the string coupling and 
 $n_g^{\bf d}$  the Gopakumar--Vafa invariants. These can be easily computed with the topological vertex \cite{akmv2} formalism or the holomorphic anomaly equation \cite{bcov}.
 For the local $\IP^1 \times \IP^1$  geometry we have for instance
 \be F^{\rm GV}({\bf{t}}, g_s)= \frac{2 q( \re^{-t_1} +  \re^{-t_2})}{(q-1)^2}+\frac{4 q  \re^{-t_1}  \re^{-t_2}}{(q-1)^2}+\frac{q^2  \re^{-2t_1}}{(q-1)^2 (q+1)^2}+\frac{q^2  \re^{-2t_2}}{(q-1)^2 (q+1)^2}+ \mathcal{O}(\re^{-3 t_i}) ,\ee
 where $q=\re^{\ri g_s}$.

 Similarly the free energy of refined topological string  in the NS limit  reads \cite{ikv}
\be \label{fns} F^{\rm NS}({\bf{t}}, \hbar)={1\over 6 \hbar} a_{ijk} t_i t_j t_k +b^{\rm NS}_i t_i \hbar + F^{\rm NS}_{\rm inst}({\bf{t}}, \hbar) -\sum_{n\geq 1}\frac{1}{n^2}{e^{-n t_2 } \cot \left(\frac{n \hbar }{2}\right)} \ee
where  
\be \ba  \label{fninst} F^{\rm NS}_{\rm inst}({\bf{t}}, \hbar)= \sum_{j_L, j_R} \sum_{w, {\bf d} } &
N^{{\bf d}}_{j_L, j_R}  \frac{\sin\frac{\hbar w}{2}(2j_L+1)\sin\frac{\hbar w}{2}(2j_R+1)}{2 w^2 \sin^3\frac{\hbar w}{2}} \re^{-w {\bf d}\cdot{\bf  t}},\\
 &  \qquad   {\bf d}=\{d_1, d_2\}, \quad d_1>0, \quad d_2 \geq 0 \ea \ee
and $N_{j_L,j_R}^{\bf d}$  denote the refined BPS invariants \cite{ckk,no}. 
These can be computed by using the 
 refined topological vertex \cite{ikv} or the  refined holomorphic anomaly \cite{hk, hkpk}.
 The last term in \eqref{fns} is often called the one-loop contribution to the NS free energy.
 For the local $\IP^1 \times \IP^1$  geometry we have for instance
 \be F^{\rm NS}({\bf{t}}, \hbar)= \frac{t_1^3}{6 \hbar }-\frac{t_1^2 (t_1-t_2)}{4   \hbar }-\frac{t_1 \hbar }{12}- \cot \left(\frac{\hbar }{2}\right) \left( \re^{-t_1}+\re^{-t_2}\right) + \mathcal{O}(\re^{-2 t_i}).\ee
The expressions \eqref{fgv}, \eqref{fns} are valid at the large radius point of the moduli space where
\be t_2, t_1 \to \infty. \ee
 However, thanks to the refined topological vertex formalism \cite{ikp,ikv},  it is possible to perform  a partial resummation in $t_2$ 
 to obtain an expression which is valid around
 \be  t_1 \to \infty, \qquad t_2 \to 0.  \ee
 As an example we consider  the standard free energy
  of local $\IP^1 \times \IP^1$. By using the refined topological vertex we obtain 
 \be \label{gvp1p1}\ba &F^{\rm GV}({\bf{t}}, g_s)=F_{\rm ol}(\re^{-t_2})+\frac{2 q \re^{-t_1}}{(q-1)^2 (\re^{-t_2}-1)^2}+\\
&\frac{q^2 \re^{-2t_1}\left(q^2 \re^{-4t_1}+q^2+4 q (q+1)^2 \re^{-3t_2}-2 (q (q+1) (q (q+3)+4)+1) \re^{-2 t_2}+4 q (q+1)^2 \re^{-t_2}\right)}{(q-1)^2 (q+1)^2 (\re^{-t_2}-1)^4 (q-\re^{-t_2})^2 (q \re^{-t_2}-1)^2}\\
& +\mathcal{O}(\re^{-3t_1}) , \qquad q=\re^{\ri g_s} ,\ea \ee
where $F_{\rm ol}(\re^{-t_2})$ is what we call the one-loop contribution of the standard topological strings. In appendix \ref{ap3} we show that, when this is appropriately combined with the one-loop contribution of the NS free energy, one can resum it by using the methods of \cite{ho2} .

Similarly, by using the refined topological vertex one has
\be \label{nsintsr} F^{\rm NS}_{\rm inst}({\bf{t}}, \hbar)=\frac{\ri  e^{\ri \hbar } \left(1+e^{\ri \hbar }\right)\re^{-t_1} }{\left(-1+e^{\ri \hbar }\right) \left(-\re^{-t_2}+e^{i \hbar }\right) \left(-1+\re^{-t_2} e^{\ri \hbar }\right)}+\mathcal{O}(\re^{-2 t_1}).  \ee

\section{Integral representation for the one-loop contribution} \label{ap3}
In this appendix we use the results of \cite{ho2}  to compute the one-loop part \eqref{f1lr} of the spectral determinant \eqref{sd2}.
It was shown in \cite{ho2}  that
\be \ba &
\sum_{m=1}^\infty \frac{(-1)^m}{ 2 m}\left( \sin \frac{m g_s}{2}\right)^{-2} \re^{-m t} -\sum_{\ell=1}^\infty \frac{1}{4\pi \ell^2}
2\csc\left( \frac{2\pi^2 \ell}{g_s}\right) \left[
\frac{2\pi \ell}{g_s}\left(t\right)+\frac{2\pi^2 \ell}{g_s} \cot\left( \frac{2\pi^2 \ell}{g_s}\right)+1 \right]
\re^{-\frac{2\pi \ell t}{g_s} }
\\& = 2\frac{\text{Li}_3(-\re^{-t})}{g_s^2}
-2\int_0^{\infty} \rd x\frac{x}{\re^{2\pi x}-1}\log(1+\re^{-2t}+2\re^{-t}\cosh g_sx). \ea \ee
With some algebraic manipulations and by following  \cite{ho2} we can write it as
\be \ba &
\sum_{m=1}^\infty \frac{1}{ 2 m}\left( \sin \frac{m g_s}{2}\right)^{-2} \re^{-m t} -\sum_{\ell=1}^\infty \frac{1}{4\pi \ell^2}
2\cot\left( \frac{2\pi^2 \ell}{g_s}\right) \left[
\frac{2\pi \ell}{g_s}t+1 \right]
\re^{-\frac{2\pi \ell t}{g_s} } -{\pi \over g_s}\sum_{\ell=1}^\infty{1\over \ell}  \cot\left( \frac{2\pi^2 \ell}{g_s}\right)^2 \re^{-\frac{2\pi \ell t}{g_s} }\\
&=-{\pi \over g_s} \log \left( 1-\re^{-\frac{2\pi  t}{g_s} }\right)+2 \frac{\text{Li}_3(\re^{-t})}{g_s^2}
- 2 {\rm Re} \int_0^{\infty\re^{\ri 0}} \rd x\frac{x}{\re^{2\pi x}-1}\log(1+\re^{-2t}-2\re^{-t}\cosh g_sx), \ea \ee
which reproduces precisely  \eqref{f1lr}. 
This means that the one-loop part of the standard topological string  plus the  one-loop part of the NS limit of topological string  sum up to give the non perturbative free energy of topological string on the resolved conifold as given in \cite{ho2}.  

\section{Spectral determinant and Painlev\'e III equation}\label{zamosd}

In this section we briefly review the results of \cite{zamo}.These results will be relevant in Section \ref{sp}.

We define the  {\it Zamolodchikov spectral determinant} $\Xi_{\rm Z}(\kappa, t)$ as
\be\label{smallzamo} \Xi_{\rm Z}(\kappa, t)=\sum_{N\geq 0} \kappa^N{D_N(t)\over N!},\ee
where
\be \label{zamomm} D_N(t)= \int  \prod_{i=1}^N {dz_i\over z_i}\re^{- 2 \sum_{i=1}^N  u(z_i, t)} \frac{\prod_{i < j} (z_i-z_j)^2}{\prod_{i <j} (z_i+z_j)^2},\ee
with 
\be \label{up}u(z,t)={t\over 4} z+{t\over 4} z^{-1}, \quad t>0.\ee
From \cite{zamo, wu1,twp3} it follows that $ \Xi_{\rm Z}(\kappa, t)$ satisfies 
\be \label{logzamo}4 \left(\left({d\over dt}\right)^2+{1\over t}{d\over dt}\right) (-\log \Xi_{\rm Z}(\kappa,t))=\left({ \Xi_{\rm Z}(-\kappa,t)\over  \Xi_{\rm Z}(\kappa,t)}\right)^2-1. \ee
Similarly \be U(\kappa,t)=2 \ri \log \left(  \ri \Xi_{\rm Z}(-\kappa,t)/ \Xi_{\rm Z}(\kappa,t)\right)\ee
satisfies 
\be \label{piii}\left(\left({d\over dt}\right)^2+{1\over t}{d\over dt}\right)   U(\kappa,t) =-\sin (  U(\kappa,t) ). \ee
In the context of  Painlev\'e  equations it useful to introduce the so called $\tau$ function which is related to the solution of \eqref{piii} as
\be \re^{-\ri U(\kappa,t) }={4 \over t }{d \over dt} t {d \over dt}\log \tau (t^4 2^{-12}, \kappa).\ee
From \eqref{logzamo} it follows that
\be \label{zamoeq}{4\over t}{d\over dt}t {d\over dt} \log \left[\Xi_{\rm Z}(\kappa,t)\re^{-t^2/16 }\right]=\left( \ri { \Xi_{\rm Z}(-\kappa,t)\over  \Xi_{\rm Z}(\kappa,t)}\right)^2.\ee
This means that \be \Xi_{\rm Z}(\kappa,t)\re^{-t^2/16 } \ee
is the $\tau$ function  corresponding to the solution of \eqref{piii}.
The small $t$ expansion of $\Xi_Z(\kappa,t)$ was also computed in \cite{zamo} where the author shows that
\be\label{zamoBG} \ba  \Xi_{\rm Z}(\kappa,t)\approx  \left({t \over 8}\right)^{4 \sigma ^2-\frac{1}{4}} &\exp \left[3 \zeta'(-1)+{5\over 6}\log2 -4 \sigma ^2 (\log (8)-1)+2 \sigma  \text{log$\Gamma $}(-2 \sigma ) \right] \\
\times&\exp \left[-2 \sigma  \text{log$\Gamma $}(2 \sigma )+\psi ^{(-2)}(-2 \sigma )+\psi ^{(-2)}(2 \sigma )\right].
\ea\ee
The variable $\sigma$ in \eqref{zamoBG} is related to $\kappa$  through 
\be \label{szamo}2\pi \kappa =\cos{2\pi \sigma }\ee
 and we can assume without loss of generalities  $ 0 \le {\rm Re}(\sigma)  \le 1/2$. 
Moreover for small values of $t$ it was shown in \cite{zamo} that
 \be \re^{\ri  U(\kappa,t)} \approx- \left({t\over 8}\right)^{8\sigma-2} {\Gamma^2\left({1}-{2\sigma }\right)\over \Gamma^2\left({2\sigma }\right)}.\ee
In particular the  monodromy data of the related Fuchsian system for this solution are
\be(\sigma, \eta=0). \ee

\bibliographystyle{JHEP}
\bibliography{biblio}
\end{document}